\documentclass[prd,twocolumn,aps,amsmath,nofootinbib,amssymb,preprintnumbers,floatfix]
{revtex4}

\usepackage{graphicx}
\usepackage{dcolumn}
\usepackage{bm}
\usepackage{amsmath}
\usepackage{amsfonts}
\usepackage{hyperref}

\def\ls{\mathrel{\lower4pt\vbox{\lineskip=0pt\baselineskip=0pt
           \hbox{$<$}\hbox{$\sim$}}}}
\def\gs{\mathrel{\lower4pt\vbox{\lineskip=0pt\baselineskip=0pt
           \hbox{$>$}\hbox{$\sim$}}}}
\def\drawbox#1#2{\hrule height#2pt
there are very few parameters and

which are well constrained by the nature of SUSY
breaking at a TEV scale.        \hbox{\vrule width#2pt height#1pt \kern#1pt
              \vrule width#2pt}
              \hrule height#2pt}

\def\Asym#1#2{\vcenter{\vbox{\drawbox{#1}{#2}
              \kern-#2pt       
              \drawbox{#1}{#2}}}}

\newcommand{\beq}{\begin{equation}}
\newcommand{\eeq}{\end{equation}}

\usepackage{epsfig,bbm,bm,amsmath,amssymb}

\def\be{\begin{equation}}
\def\ee{\end{equation}}    
\def\baray{\begin{eqnarray}}
\def\earay{\end{eqnarray}}

\def\2pi{\left(2\pi\right)}

\begin{document}

\title{Exciting gauge field and gravitons in a brane-anti-brane annihilation}

\author{Anupam Mazumdar $^{1,2}$}
\author{Horace Stoica $^{3}$}

\affiliation{$^{1}$~Physics Department, Lancaster University, LA1 4YB, UK\\
$^{2}$~Niels Bohr Institute, Copenhagen University, Blegdamsvej-17,
  DK-2100\\
$^{3}$~Theoretical Physics, Blackett Laboratory, Imperial College, London, 
SW7 2AZ, UK.
}

\begin{abstract}
In this paper we point out the inevitability of an explosive production of gauge field and gravity 
wave during an open string tachyon condensation in a cosmological setting, in an effective field theory model. 
We will be particularly studying a toy model of brane-anti-brane inflation in a warped throat where inflation ends via tachyon 
condensation. We point out that a tachyonic instability helps fragmenting the homogeneous tachyon 
and excites gauge field and contributes to the stress energy tensor which also feeds into the gravity waves. 
\end{abstract}

\maketitle

There are many cosmological sources of gravity wave generation, for a review see~\cite{Rev}. In 
this paper we will show how an interesting scenario emerges in string cosmology where there 
exists plethora of examples of open string tachyon condensation~\cite{Sen}. 
We will show how gravity waves and gauge field are excited in a field theory model of brane-anti-brane 
annihilation particularly in a cosmological setting.

The tachyon plays an important role in terminating brane-anti-brane inflation~\cite{Burgess}, 
for a review see~\cite{Rev1}, 
when a pair comes close to each other, less than a critical distance. When the branes are far apart the 
tachyon field does not play any role, while the brane-anti-brane separation plays the role of an 
inflaton and the brane tension generate the vacuum energy density. The slow attraction of the 
branes towards each other gives rise to a primordial inflation and also a graceful exit. In this 
process the so called tachyon field acts almost like a waterfall field like in the case of a 
hybrid inflation~\cite{Hybrid}, where the $({\bf mass})^2$ of the waterfall field changes from 
being positive-to-zero-to-negative. In all these examples the waterfall field and the tachyon 
in string theory examples are considered to be a homogeneous field evolving with time.

In this paper we discuss couple of important consequences of a brane-anti-brane 
inflation~\footnote{Although end of inflation via brane-anti-brane lends the greatest 
support for studying the tachyons, but open string tachyons could also give rise to primordial 
inflation~\cite{AM}, and furthermore tachyons can possibly generate large non-Gaussianity~\cite{AM1}. }
in the field theory model. Note that field
theoretical treatment is an oversimplification of the brane annihilation process,
but we will be content with it as a toy model to investigate excitation of gravity
wave and gauge field. It would be interesting to see how
stringy effects will modify our findings.

First, in a cosmological context the tachyon will never keep the homogeneity, the tachyon 
will be fragmented by virtue of inhomogeneity created by the quantum fluctuations, quite similar 
to the tachyonic preheating~\cite{Linde}. Second, the tachyon in string theory couples bi-linearly 
to the gauge fields residing on the world volume of the branes. It can be shown by field 
redefinitions that in an Einstein frame the gauge fields couple minimally, therefore, the 
gauge fields are {\it Higgsed} by the tachyon, i.e. the gauge fields obtain v.e.v. dependent 
masses. This is an ideal case when a time dependent  v.e.v.  can excite gauge fields at 
sub-and super-Hubble  scales.

Third, the fragmentation of the tachyon and its inhomogeneous perturbations seeds 
anisotropic stress tensor which leads to gravity wave generation. 
The frequency of these gravity waves depends on the string scale which we take it to be close to
$M_s\sim 10^{14}$~GeV (below the scale of grand unified theory). Since the scale is quite
high therefore the frequency will not be detectable by future gravity wave experiments. 

In particularly, we will be studying inflation in a warped throat,
while the SM or the MSSM fields are located in a different throat with
a different warping. For simplicity we will consider a simple
$D3-\overline{D3}$ system, where the modulus determining the inter
brane separation  drives inflation and inflation ends via tachyon
condensation~\footnote{Inflation in such a setup will not directly
reheat the MSSM throat~\cite{Kofman}, reheating happens indirectly via
moduli belonging to the MSSM~\cite{Rob}, see also~\cite{Robert}. One plausible solution is
to inflate the universe within the MSSM throat ~\cite{MF},
which would naturally reheat the universe with the MSSM  degrees of freedom.}. However
note that the branes do carry gauge fields in their worldvolume, the tachyon in 
a brane-anti-brane system is a bi-fundamental field that couples only to a linear 
combination of the two gauge fields. These gauge fields drop out of the dynamics 
once the pair has annihilated~\footnote{ However, more realistic models would 
include a stack of $N$ (anti)branes located at the bottom of the throat and a 
number of $M$ mobile branes moving towards them. After 
annihilation the remaining stack of branes supports an unbroken  
$U\left(N-M\right)$ \cite{Horava:1998jy,Witten:1998cd,Sarangi:2002yt}. The original  
$U\left(N\right)\times U\left(M\right)$ is higgsed  down to $U\left(N-M\right)$ when 
the tachyon develops a v.e.v. at the end of inflation.}. 

When the brane and the anti-brane are coincident the action 
for the tachyon field responsible for the anniliation of the
pair is described by \cite{Kraus:2000nj,Takayanagi:2000rz}:
\baray
&&{\mathcal S}=-2T_{D3}\int \sqrt{-g_{4}} d^{4}x \,
e^{-2\pi\alpha^{\prime}T\overline{T}}\biggl[
1+\biggr.
\nonumber \\
&& \biggl. 
8\pi\left(\alpha^{\prime}\right)^{2}\ln\left(2\right)D^{\mu}\overline{T}D_{\mu}T
+ 
\frac{\gamma{\alpha^{\prime}}^{2}}{8}
\left(F^{+}_{\mu\nu}-F^{-}_{\mu\nu}\right)^2\biggr]
\label{BaB}
\earay
Here $F^{+}$ and $F^{-}$ are the gauge fields that live
in the worldvolume of the brane and anti-brane respectively.
The tachyon is a bi-fundamental field that couples only 
to a linear combination of the two gauge fields 
$D_{\mu}T = \partial_{\mu}T - \left(A_{\mu}^{+}-A_{\mu}^{-}\right)T$.
If we expand the exponential term as:
$e^{-2\pi\alpha^{\prime}T\overline{T}} = 1 - 2\pi\alpha^{\prime}T\overline{T} $ 
and write the $D3$ brane tension as: $
T_{D3} = {1}/{(\left(2\pi\right)^{3}\left(\alpha^{\prime}\right)^{2}g_{s})}$.
The numerical pre-factor $8\pi\ln\left(2\right)$ can be absorbed in a field
re-definition and the action  with $\chi = (\sqrt{2\ln 2}~r /{\pi R})T$ 
can be brought to  the form (the scalar part only)~
\footnote{
To find the warped tachyon mass we write the metric in the 
warped throat takes the form (up to corrections in 
a KS type throat): $ds_{10}^{2}=\frac{r^2}{R^2}ds_{4}^{2} + 
\frac{R^2}{r^2}\left(dr^2 + r^{2}d\Omega_{5}^{2}\right)$, see Ref.~\cite{Cremades:2005ir}.}
:
\baray
S&=& \int d^4x \left(\frac{2r^{4}}{R^{4}}T_{D3} + 
\frac{1}{2}D_{\mu}\chi D^{\mu}\chi - 
\frac{r^{2}}{R^{2}}\frac{M_{s}^{2}}{2\ln\left(2\right)}\chi^2\right) 
\nonumber \\
&=& \int d^4x \left(\Lambda_{4} + 
\frac{1}{2}D_{\mu}\chi D^{\mu}\chi - 
\frac{1}{2}m_{T}^{2}\chi^{2}\right)
\label{unstable_brane}
\earay
Denoting $r/R = w$, the warp factor, we see that the cosmological constant 
given by the brane tension and the tachyon mass depend only on the warped
string scale $m = wM_{s}$, and therefore the Hubble constant during inflation 
and the tachyon mass during reheating are related $
\Lambda_{4} = 2w^4T_{D3} = w^4{2M_{s}^{4}}/{\left(2\pi\right)^{3}g_{s}} = 
{2m^4}/{\left(2\pi\right)^{3}g_{s}}$

\begin{figure}[htbp]
\begin{center}
\includegraphics[width=0.25\textwidth,angle=0]{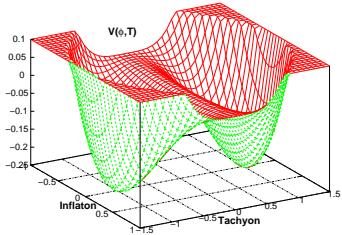}
\end{center}
\caption{The polynomial approximation we use has the advantage of 
  allowing us to modify the tachyon mass at $\phi = 0$ 
  independently of the quartic term. The downside is that the 
  the value of the potential at its minimum is negative. 
  \label{tachyon_pot}}
\end{figure}


In order to understand the aftermath of inflation, we need to include the gauge fields and we also
need to model the scalar potential in the tachyonic phase by a polynomial function of the fields. 
\baray
&& {\mathcal L} = \frac{1}{2}\partial_{\mu}\phi\partial^{\mu}\phi +
\frac{1}{2}D_{\mu}\chi D^{\mu}\chi^{*} +
\frac{1}{2}m_{\text{inflaton}}^{2}\phi^{2}
\nonumber \\
&& +\frac{1}{2}m_{T}^{2}\left|\chi\right|^{2}\left(\phi^{2}-\phi_{0}^{2}\right)+
\frac{\lambda}{4}\left|\chi\right|^{4}
-\frac{1}{4}F^{2}
\label{approx_L}
\earay

The value of $\phi_{0}$ represents the critical distance at which the
tachyon field becomes unstable. We had to add a quartic term for the 
tachyon potential, so that the overall potential is bounded from below. 
Otherwise the simulation cannot be performed. 
We are interested in the dynamics of the tachyon at the top of its 
potential, $\chi \approx 0$, where the quartic term is negligible.

We have verified that the quartic term does not change the results
of our simulations as far as the tachyonic phase is concerned. 
Making the quartic term smaller will result in a potential with 
deeper minima, and therefore a longer tachyonic phase. However, 
since the VEV of the field increases exponentially with time, the 
duration of the tachyonic phase depends only logarithmically in the
coefficient of the quartic term. The quartic term has a large 
influence during the turbulent phase after the tachyonic amplification 
has ended. Since the minimum is deeper, the field will reach this 
minimum with higher kinetic energy, and therefore more energy is 
transferred to the gravity waves in a given time interval.

The Lagrangian in Eq.(\ref{BaB}) describes the brane-anti-brane system 
only when the brane and the anti-brane are coincident. In our approximate 
Lagrangian, Eq.(\ref{approx_L}), we include the dependence of the tachyon 
mass on the inflaton v.e.v., that is the brane-anti-brane separation.

To study the gravity waves, we follow the approach of \cite{GarciaBellido:2007af}
were the perturbations of the Einstein equations are studied. We start by considering
the metric fluctuations: $ds^{2}=-dt^{2}+a^{2}\left(t\right)\left(\delta_{ij}+h_{ij}\right)dx^{i}dx^{j} $.
The corresponding equation for the fluctuations of the metric reads: $
\Box h_{ij} = 16\pi G \delta T_{ij}$,
where $T_{ij}$ is the anisotropic stress-energy tensor. Just as in \cite{GarciaBellido:2007af}
we follow the evolution of only the traceless part of the tensor fluctuations:
$\ddot{h_{ij}}-\nabla^{2}h_{ij}=16\pi G\Pi_{ij}$, where $\Pi_{ij} = T_{ij}-\frac{1}{3}T\delta_{ij}$.
We may neglect the expansion of the universe since $M_{s} \gg H$ at the time when tachyon is rolling
after inflation.

The stress-energy tensor gets contributions from  both the charged tachyon and uncharged 
inflaton, as well as the gauge fields. Details of the calculation of the stress-energy tensor will 
be presented in a separate paper. Here we simply quote the results:
\baray
\Pi_{ij} &=& F_{iC}F_{j}^{\;\;C} - 
\frac{1}{3}\delta_{ij}F_{kC}F^{kC} 
- D_{i}\chi D_{j}\chi^{*} \nonumber \\
&+& \frac{1}{3}\delta_{ij}D_{k}\chi D^{k}\chi^{*}
 -\partial_{i}\phi \partial_{j}\phi^{*} +  
\frac{1}{3}\delta_{ij}\partial_{k}\phi \partial^{k}\phi^{*}
\earay
When studying the effects of the gauge fields
we look at both the energy the gauge fields pick up during the 
evolution as well as their contribution as a source for the 
gravitational waves. The energy density of the gravitational 
waves is simply the $t_{00}$ component of the 
energy-momentum tensor of the gravitational waves: 
$
t_{00}=\frac{1}{32\pi G}\left<\partial_{0}h_{ij}^{TT}\partial_{0}h^{ij}_{TT}\right>$ 
where $TT$ stands for the transverse-traceless part of the tensor fluctuations. 
The energy density calculated simply for $h_{ij}$ is related to that for 
$h_{ij}^{TT}$ by a numberical factor given by the angular integral of the 
projection operator that trasforms $h_{ij}$ into $h_{ij}^{TT}$.  We also 
use the synchronous gauge for the metric, so the 
$h_{00}$ and the $h_{0i}$ components of the metric fluctuations 
are set to zero. 

\begin{figure}[tbph]
\begin{center}
\includegraphics[width=0.30\textwidth,angle=0]{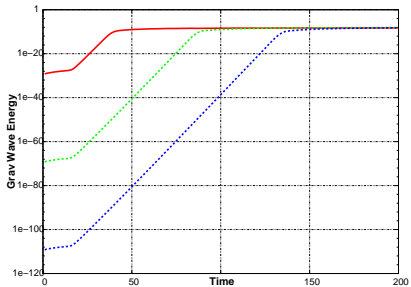}
\vspace{6pt}
\end{center}
\caption{Gravity wave energy with varying initial amplitudes. Note that different
initial amplitudes saturate at a similar value for a fixed tachyon mass $\sim M_{s}/2$.
  \label{grav-1}}
\end{figure}

\begin{figure}[tbph]
\begin{center}
\includegraphics[width=0.30\textwidth,angle=0]{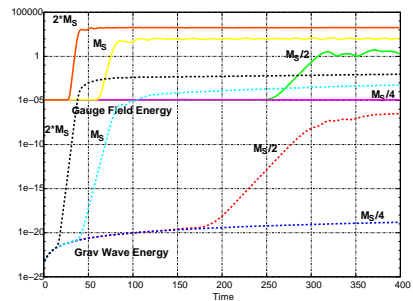}
\end{center}
\caption{The energy pumped into the gauge field and the gravity waves.
\label{GG}}
\end{figure}

The parameters of the model are set to reflect the typical values for an 
inflationary model. We set the charge coupling the gauge field to the 
tachyon to $1$, and the same for the quartic self-coupling of the tachyon, 
$\lambda=1$. The mass of the inflaton is taken to be $m_{\phi}^{2} = 0.01$, 
much smaller than the tachyon mass. We set the Newton constant to a value 
corresponding to the typical string mass we obtain by requiring that the 
brane-anti-brane inflationary model reproduces the observed amplitude of 
the CMB fluctuations, $M_{S}/M_{P}=10^{-4}$~\cite{Rev1,Sarangi:2002yt,Battye:2007si}. This 
value for the ratio 
between the string and Planck scales tells us that all the modes in our
simulation are inside the Hubble horizon. The Hubble constant during, and 
shortly after inflation is: $
H^{2} = {V}/{3M_{P}^{2}}\simeq {M_S^4}/{3M_{P}^{2}} = 
 ({M_{S}}/{M_{P}})^{2}M_{S}^{2}/3 $.
We can now see how the Hubble radius compares with the size of our 
simulation box $\sim 30 {\it l}_{S}$, taking into account that 
the lattice spacing is about half the string length ${\it l}_{S} = 1/M_{S}$:
$\frac{1}{H} \simeq \frac{M_{P}}{M_{S}}{\it l}_{S} = 
10^{4}{\it l}_{S} \gg 30 {\it l}_{S}$.
The value of the tachyon mass, controlled by the coupling of the tachyon to 
the inflaton field, is chosen to take $4$ values and we compare the production 
of gravitational waves when the tachyon mass is changed. The time scale for
the instability to grow depends on inversely proportional to the mass scale 
of the tachyon, therefore, a larger massive tachyon leads to a quicker exponential
growth in gauge field and in gravitational waves compared to the smaller one,
see Fig.~(\ref{GG}). 

\begin{figure}[tbph]
\begin{center}
\includegraphics[width=0.30\textwidth,angle=0]{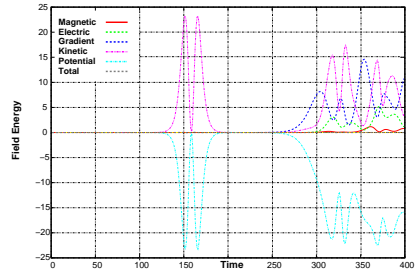}
\end{center}
\caption{The energy components  of the tachyon and gauge field for the 
tachyon mass $M_{S}/2$. Note that the total energy remains conserved.
  \label{tachyonic}}
\end{figure}

For the initial condition, we assume that the tachyon mass is effectively vanishing towards 
the end of inflation where the inflaton v.e.v. is $\approx \phi_0$. We then include initial 
fluctuation of the tachyon around this v.e.v. with a spectrum given by the fluctuations
of a scalar in the deSitter phase: $\chi\left(k\right)={c_{k}}/{k^{3/2}}$. 
The Fourier coefficients of the field fluctuations had been taken in 
\cite{GarciaBellido:2007af} to follow a Gaussian distribution with a 
dispersion dependent on the wave number as $\sigma_{k}^{2}\sim k^{-3}$.
Here we will approximate the Gaussian distrbution with a uniform one 
in the interval $\left(-{1}/{2k^{3/2}}, {1}/{2k^{3/2}}\right)$.
We obtain this distribution by choosing $c_{k}$ uniformly distributed 
in the interval $\left(-{1}/{2}, {1}/{2}\right)$ and then using 
${c_{k}}/{k^{3/2}}$ as the Fourier coefficients for the tachyon, thus
the real and imaginary parts are being independent random numbers.

\begin{figure}[tbph]
\includegraphics[width=0.11\textwidth,angle=0]{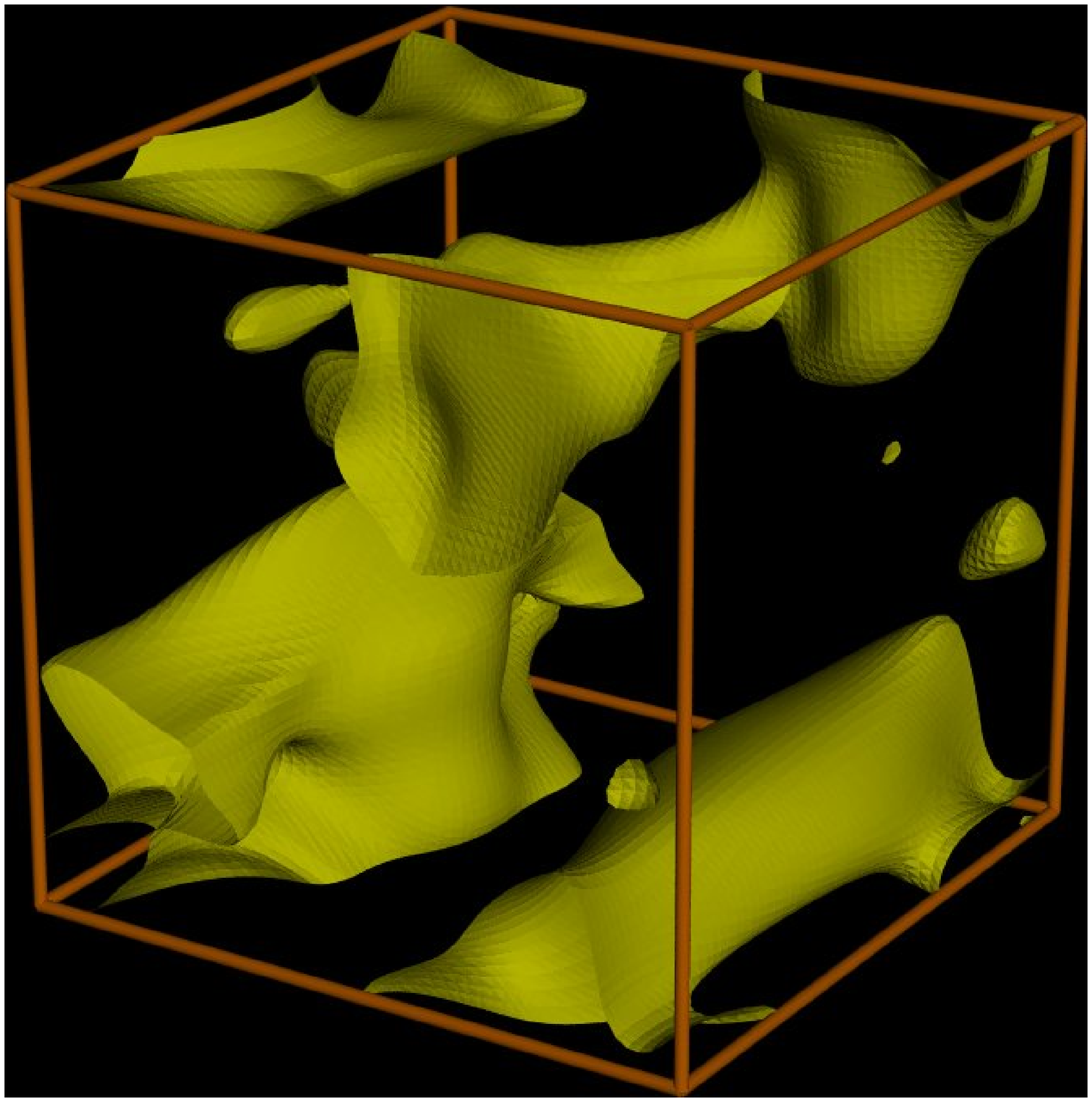}
\includegraphics[width=0.11\textwidth,angle=0]{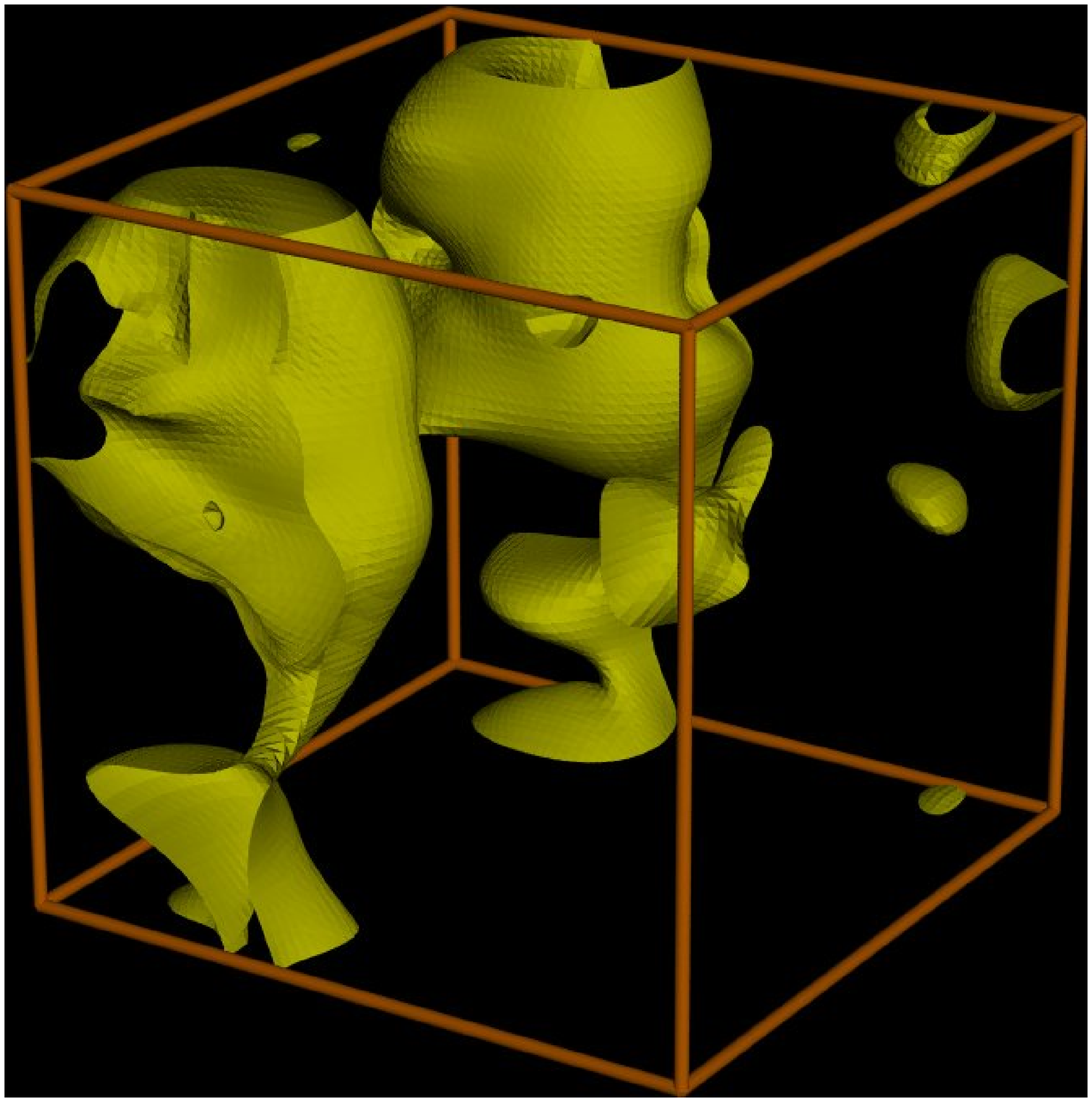}
\includegraphics[width=0.11\textwidth,angle=0]{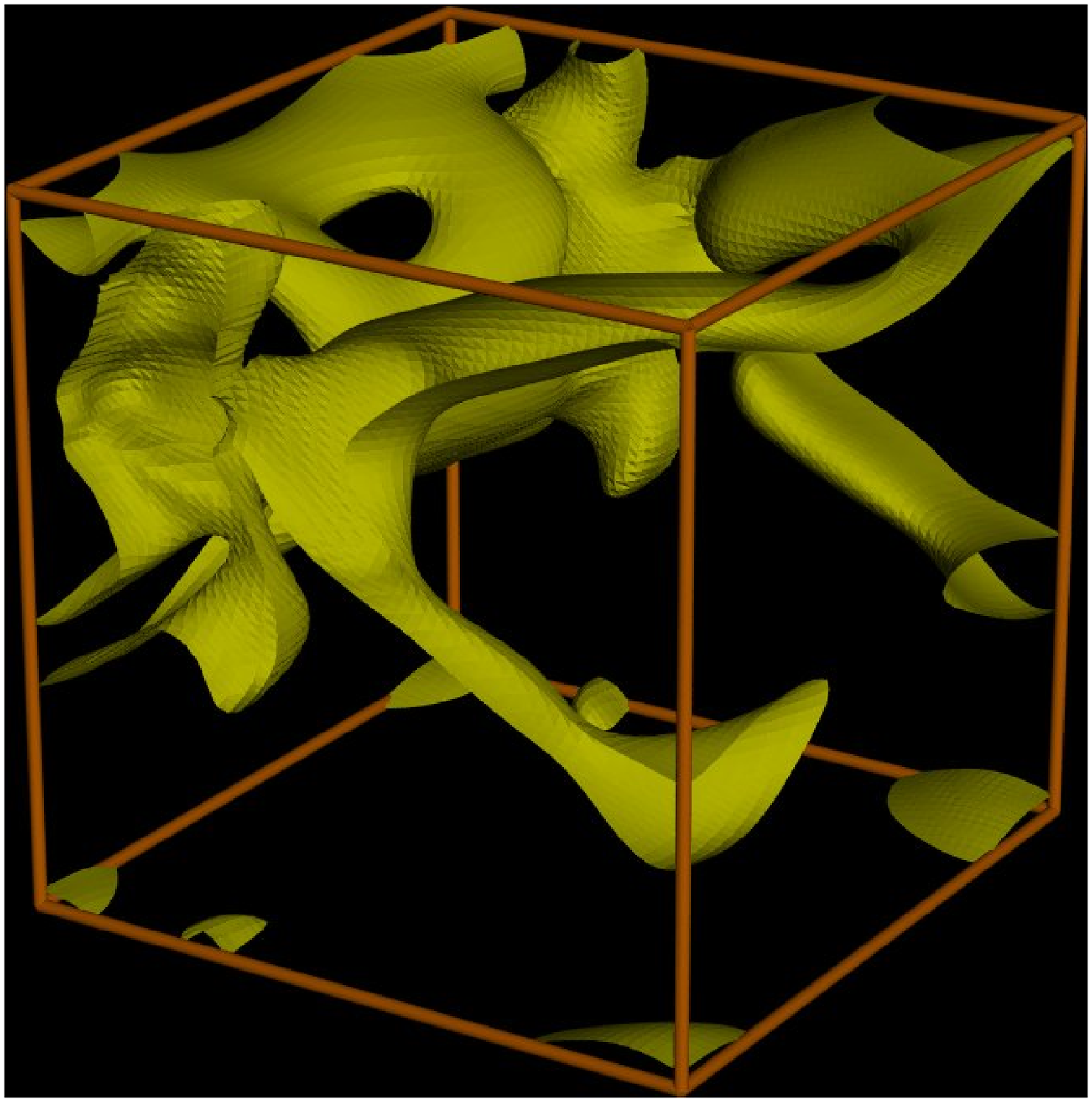}
\includegraphics[width=0.11\textwidth,angle=0]{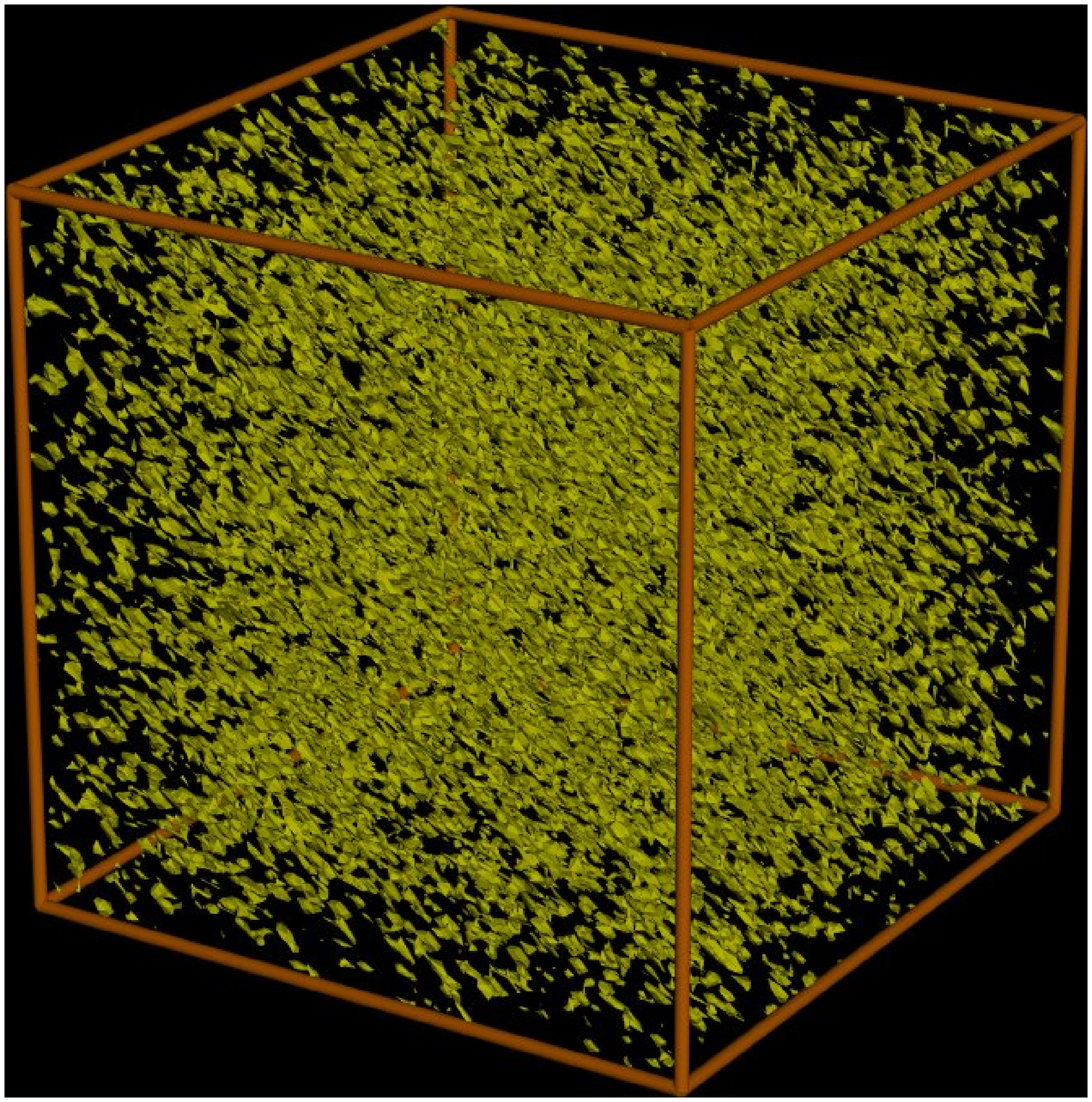}
\vspace{6pt}
\caption{Snap-shots of iso-surface of the energy density for the gauge field, gravity 
waves, tachyon  and the inflaton (from left-to-right) at $t=300$.
  \label{tachyonic-1}}
\end{figure}

The inflaton starts with a v.e.v. of $\left|\phi\right| = \phi_0$, corresponding 
the the onset of the tachyonic instability. On top of this v.e.v. we 
superimpose fluctuations corresponding to the deSitter phase, 
just like for the tachyon. The gauge fields have only the fluctuations 
corresponding to the deSitter phase.  We take the gravitational waves to 
start at varying amplitude, see~Fig.~(\ref{grav-1}), all of them saturate to a similar 
amplitude. The time is measured in a string unit.

In Fig.~(\ref{tachyonic}) we have shown various energy components of the tachyon.
The potential energy of the tachyon is negative given our choice 
 of a potential. We observe that the production of gravitational waves 
 does not start until the gradient energy (dark blue curve) of the 
 tachyon starts to pick up, see Figs.~(\ref{tachyonic},~\ref{GG}). The first 
 "spike'' in the energy graph, see Fig.~(\ref{tachyonic}), corresponds to the 
 first oscillation of the tachyon, but the field 
 stays almost homogeneous during this first oscillation. When the 
 gradient energy picks up and the field condensate fragments, then only the gravity 
 waves are produced and also the gauge field is excited.  The energy in the $\phi$ field 
 remains negligible.

The energy stored in the gauge field also get exponentially amplified 
due to the fragmentation of the tachyon field.
Although they carry most of the energy, see Fig.~(\ref{GG}), compared to the gravity waves,
their amplification happens at a slightly later stage, i.e. compare the two in Fig.({\ref{GG})
for tachyon mass $M_s/2$.

In Figs.~(\ref{tachyonic-1}) we have shown 4 snap-shots of the iso-surface of the constant 
energy density for the gauge field, the gravity waves, the tachyon and the inflaton. Couple of 
points to note here, all the fields except the inflaton shows a remarkable departure in the 
homogeneity. Except the inflaton all fields undergo long wavelength excitations (they all 
look relatively smooth on small scales), while the inflaton obtains the largest inhomogeneity 
on the smaller scales. This is due to the fact that the there is no long wavelength amplification 
for the inflaton, rather the inflaton starts oscillating rapidly with an enhanced frequency by virtue 
of the coupling  $\sim m^2_{T}|{\chi}|^2$. However as the tachyon fragments, the $\chi$ field no 
longer retains the homogeneity, therefore the inflaton also obtains space dependent mass which 
fragments the inflaton on sub-Hubble scales.

Let us now comment on the gravity waves, its amplitude and frequency. Due to the tachyonic 
growth the amplitude of the frequency is given by: 
$\rho_{GW}/\rho_0 \sim t_{00}/M_S^4\sim \langle \dot h_{ij}\dot h^{ij}\rangle_V$, where $V$
stands for the volume average. From Fg.~(\ref{GG}) it is clear that the factor 
$\rho_{GW}/\rho_0\sim 10^{-6}$ while its frequency depends on the details of the expansion 
history of the universe, however no matter what happens lately, the frequency will be well beyond 
the reach of the future gravity wave detectors. 

To summarize, we have shown a brane-anti-brane annihilation inevitably ends up in fragmentation
of a tachyon which excites gauge field and gravity waves simultaneously. We also note that the
excitations are generically long wavelength in nature, except that of the inflaton. However our 
lattice box size is such that we can only explore sub-Hubble processes.The amplification of 
gauge and gravity waves depends on the mass of a tachyon. A larger mass tachyon leads 
to a quicker growth as compared to a smaller tachyon mass. The amplitude of the gravity 
waves could be quite large, i.e. $\rho_{GW}/\rho_0\sim 10^{-6}$ but their frequency will be 
towards the high end of the spectrum where future detection will not be foreseeable.
Our simulation also unveils the first study of exciting gravity waves where the  gauge field 
has been taken into account. We found that the initial burst of gravity waves arises from the fragmentation
of the tachyon field, at only later stages the gauge field feeds into the production of gravity waves.
One point to note here that we have not studied turbulence of the gauge field and the gravity waves, which will
be considered in a separate publication.

{\it Acknowledgements:-} The numerical simulations have been done on the 
Imperial College HPC cluster. AM is partly supported by grant (MRTN-CT-2006-035863).

\vspace{15pt}



\end{document}